\documentclass[12pt,preprint]{aastex}

\slugcomment{To appear in {\it The Astrophysical Journal}}
\received{}
\accepted{}
\journalid{}{}
\articleid{}{}
\begin{document}

\title{Search for a Point-Source Counterpart of the Unidentified Gamma-Ray
Source TeV J2032+4130 in Cygnus}

\author {R. Mukherjee\altaffilmark{1}, J. P. Halpern\altaffilmark{2,3},
E. V. Gotthelf\altaffilmark{3}, M. Eracleous\altaffilmark{4,5},
N. Mirabal\altaffilmark{2}}

\altaffiltext{1}
{Department of Physics \& Astronomy, Barnard College, New York, NY 10027}
\altaffiltext{2}
{Department of Astronomy, Columbia University, New York, NY 10027}
\altaffiltext{3}
{Columbia Astrophysics Laboratory, Columbia University, New York, NY 10027}
\altaffiltext{4}
{Department of Astronomy \& Astrophysics, The Pennsylvania State University,
University Park, PA 16802}
\altaffiltext{5}
{Visiting Astronomer, Kitt Peak National Observatory,
National Optical Astronomy Observatories, which is
operated by the Association of Universities for
Research in Astronomy, Inc. (AURA) under cooperative
agreement with the National Science Foundation.}

\begin{abstract}

We have made a multiwavelength study of the overlapping error boxes of the unidentified
$\gamma$-ray sources TeV J2032+4130 and 3EG J2033+4118 in the direction
of the Cygnus OB2 association ($d = 1.7$ kpc) in order to search for a
point-source counterpart of the first unidentified TeV source.
Optical identifications and spectroscopic classifications for the brighter
X-ray sources in {\it ROSAT\/} PSPC and {\it Chandra} ACIS images are obtained,
without finding a compelling counterpart.  The classified X-ray sources
are a mix of early and late-type stars, with one exception.  The brightest source
in the {\it Chandra} observation is a new, hard absorbed source that
is both transient and rapidly variable.  It lies $7^{\prime}$
from the centroid of the TeV emission, which places it outside of the
claimed $2\sigma$ location ($r \approx 4.\!^{\prime}8$). A possible eclipse or
``dip'' transition is seen in its light curve.  With a peak 1--10~keV luminosity of
$\approx 7 \times 10^{32}\,(d/1.7\,{\rm kpc})^2$ ergs s$^{-1}$,
this source could be a quiescent low-mass X-ray binary that lies beyond
the Cyg OB2 association.  A coincident, reddened optical object of
$R = 20.4,\ J = 15.4,\ H = 14.2$, and $K = 13.4$ is observed,
but not yet classified due to the lack of obvious emission or absorption features
in its spectrum.  Alternatively, this {\it Chandra} and optical
source might be a considered a candidate for a ``proton blazar,''
a long hypothesized type of radio-weak $\gamma$-ray source.
More detailed observations will be
needed to determine the nature of this variable X-ray source, and to assess the
possibility of its connection with TeV J2032+4130.

\end{abstract}

\newpage

\section{Introduction}

The serendipitous detection of emission coming from the direction of
the Cygnus OB2 stellar association by the HEGRA CT-System at
La Palma (Daum et al. 1997) represents the first discovery of an
unidentified TeV source, extending the mystery of the elusive
$\gamma$-ray sources to the TeV regime.
TeV J2032+4130 was discovered 
(Aharonian et al. 2002a) in observations originally devoted to
Cygnus X--3 and the unidentified EGRET (Energetic Gamma-ray
Experiment Telescope) source GeV J2035+4214
(Lamb \& Macomb 1997). An analysis of these combined fields led
to the detection of a source at (J2000)
$20^{\rm h}32^{\rm m}07^{\rm s}\pm 9.\!^{\rm s}2_{\rm stat}\pm2.\!^{\rm s}2_{\rm sys},
\ +41^{\circ}30^{\prime}30^{\prime\prime}
\pm 2.\!^{\prime}0_{\rm stat} \pm 0.\!^{\prime}4_{\rm sys}$.
Its error circle overlaps the edge
of the 95\% confidence error ellipse of another EGRET source, 3EG J2033+4118,
and is $\approx 0.\!^{\circ}5$ north of Cyg X--3.
It is not clear if TeV J2032+4130 is associated
with 3EG J2033+4118, which is itself of uncertain origin.
There is some evidence that the TeV emission is
extended, with a Gaussian $1\sigma$
radius of $\sim 5.\!^{\prime}6 \pm 1.\!^{\prime}7$.
Unlike the flaring blazars that have been detected so far,
TeV J2032+4130 was found to be steady in repeated HEGRA observations
from 1999 to 2001.  The integrated flux measured above 1~TeV was found to be 
$(4.5\pm 1.3_{\rm stat})\times10^{-13}$ photons cm$^{-2}$ s$^{-1}$,
which is $\approx 2.6\%$ of the flux of the Crab Nebula
(Aharonian et al. 2002a).

The majority of the
$\gamma$-ray sources detected above 100 MeV by the EGRET instrument
on the {\it Compton Gamma Ray Observatory}
(CGRO) are unidentified, some remaining so since the first surveys of the
$\gamma$-ray sky with the COS--B satellite. Resolving the nature of these
mysterious $\gamma$-ray sources is a challenge across all wavelengths.
Their relatively large error boxes make
counterpart searches difficult.  Unidentified EGRET sources typically have
positional uncertainties of
$\sim 0.\!^\circ5-1^\circ$, making identification on the basis of
position alone nearly impossible, especially in the Galactic plane.
We have found that multiwavelength studies of
EGRET fields on a case-by-case basis are often useful in finding a likely
identification. Such an approach has been
used to suggest counterparts for the EGRET unidentified sources
2EG J0635+0521 (Kaaret et al. 1999),
3EG J2227+6122 (Halpern et al. 2001a), 
3EG J1835+5918 (Halpern et al. 2002; Mirabal et al. 2000,2001),
3EG J2006--2321 (Wallace et al. 2002),
3EG J2016+3657 (Mukherjee et al. 2000; Halpern et al. 2001b),
3EG J1621+8203 (Mukherjee et al. 2002),
and 3EG 2021+3716 (Roberts et al. 2002), to name a few
examples (see the review by Caraveo 2002).

Imaging atmospheric Cherenkov telescopes (IACTs)
like HEGRA have the advantage of much better angular resolution, and
therefore smaller error boxes for $\gamma$-ray source positions in
comparison to EGRET.
In this paper, we adopt for the moment the hypothesis that TeV J2032+4130
is {\it not\/} significantly extended, in which case one should search for
and evaluate candidate point-source counterparts at other wavelengths.
Accordingly, we studied the EGRET source 3EG J2033+4118, archival radio
data from the NRAO-VLA Sky Survey (NVSS: Condon et al. 1998),
and pointed {\it ROSAT\/} X-ray observations of the field of 
3EG J2033+4118 and TeV J2032+4130.  We also examined a recent
{\it Chandra} observation that was centered on the TeV source,
and obtained optical identifications of the brightest
X-ray sources in this region using the MDM Observatory,
Kitt Peak National Observatory (KPNO),
and the {\it Hobby-Eberly Telescope} ({\it HET\/}).

\section{Gamma-ray Observations of 3EG J2033+4118}

The EGRET source 3EG J2033+4118 is located at
$l=80.\!^{\circ}27$, $b=+0.\!^{\circ}73$ (Hartman et al. 1999),
with a 95\% error ellipse defined by semi-major and semi-minor axes
of $0^\circ\!.31$ and $0^\circ\!.25$,
respectively (Mattox, Hartman, \& Reimer 2001). Figure~1 shows the
95\% confidence EGRET ellipse superimposed on a {\it ROSAT\/} X-ray image
that is described further in \S 3. 3EG J2033+4118 was detected by EGRET in
several viewing periods, with the most significant detection in 1993 March (VP 212).
Figure~2 shows the light curve measured by EGRET during 1991--1997.
The points before 1996 come from
the Third EGRET Catalog (3EG, Hartman et al. 1999),
in which the summed exposure has a flux above 100 MeV of
$(73.0\pm6.7) \times 10^{-8}$ photons cm$^{-2}$ s$^{-1}$.
The corresponding background-subtracted spectrum of
3EG J2033+4118 is hard, with a photon spectral index of $1.96\pm0.10$
(Hartman et al. 1999).

This location was also in the EGRET field of view four times after the
3EG catalog.  For two of these the source was almost $30^{\circ}$
from the axis, so we exclude them.  The remaining two detections
are $2.4\sigma$ and $4.2\sigma$.  Their fluxes are shown as the
last two points in Figure~2.
A chi-square analysis was used to calculate the ``variability index''
defined by McLaughlin et al. (1996). This index is
sometimes used to judge variability in EGRET sources, although it is
somewhat arbitrary. The variability index of 3EG J2033+4118 was found to be $V=1.4$.
For comparison, McLaughlin et al. (1996) interpret $V < 0.5$ as
non-variable and $V \geq 1.0$ as variable.

\section{X-ray and Radio Observations}

Figure~1 shows an X-ray image taken with the {\it ROSAT\/}
Position Sensitive Proportional Counter (PSPC)
in the energy range 0.2--2.0 keV, covering the field of
3EG J2033+4118/TeV J2032+4130. This image was created by co-adding
exposure-corrected skymaps of 23.5 ks of data taken during 1991 and 1993.
These fields were targeted originally to include
the Cyg OB2 association and also Cyg X-3.
The EGRET 95\% error ellipse for 3EG J2033+4118 is superposed on the image,
as is the $1\sigma$ error contour of the TeV J2032+4130 centroid position.
The larger circle around
the TeV position indicates the possible Gaussian $1\sigma$ 
extent of the TeV source (Aharonian et al. 2002a).
Figure~3 shows a {\it ROSAT\/} HRI image,
covering roughly the same field as that shown in Figure~1.
The image was created by co-adding exposure
corrected skymaps of $\approx 155$ ks of data taken during
observations of Cyg OB2 and Cygnus X-3 in 1993--1995.
The sources numbered in Figure~3 correspond to those in
Figure~1 and Table~1.

A detailed description of {\it ROSAT\/} results on the
sources in the Cyg OB2 association is given by Waldron
et al. (1998).   Several of the stars in the Cyg OB2
association are among the strongest stellar
X-ray sources in the Galaxy.
In Table~1 we list the coordinates of the 
brightest ROSAT sources as marked in Figures 1 and 3,
along with optical identifications and positions.
We have concentrated on obtaining optical identifications of
several X-ray sources that are within the region of maximum
likelihood for a point source of TeV emission.  We find that
all of these have ordinary stellar counterparts.  Further details on their
optical properties are presented in \S 4.2.

On 2002 August 11, {\it Chandra} made a 5 ks
director's discretionary observation (Butt et al. 2003) 
of the field of TeV J2032+4130
with the front-illuminated, imaging CCD array of
the Advanced CCD Imaging Spectrometer
(ACIS-I).  Several X-ray sources near the centroid
of the TeV source were detected.
Figure~4 shows the {\it Chandra} image with the brightest point
sources marked, which are those having at least 10 photons.
Their positions and count rates are listed in Table~2.
Cross-references to sources detected by both {\it ROSAT\/} and {\it Chandra}
are indicated in Tables 1 and 2.  The brightest {\it Chandra} source 
\#2 is notable in that it was {\it not\/} detected in any of the
{\it ROSAT\/} or {\it Einstein} observations of this field.
More details about this source are given in \S 5.1.

Table~3 is a list of eight 1.4 GHz sources from the NVSS within $10^{\prime}$
of the TeV centroid.  The brightest is an extended source
with flux density of only 18 mJy, which is at least an order of
magnitude fainter than the weakest candidate
radio sources for EGRET blazars
considered by Mattox et al. (1999).  Furthermore,
none of the radio sources is coincident with an X-ray source.  Thus,
it is unlikely that TeV J2032+4130 has an ordinary blazar counterpart,
although new types of blazars can be hypothesized (see \S 5.3).

\section{Optical Observations}

\subsection{Optical Imaging}

In preparation for the identification of
X-ray sources in the {\it Chandra} observation, on 2002 July~7
we obtained a deep $R$-band image of the field of TeV J2032+4130
using the MDM 1.3m telescope and a thinned, back-illuminated $2048 \times 2048$
pixel SITe CCD.  This $17^\prime \times 17^\prime$ image
covers all of the {\it Chandra} source positions marked in Figure~4.
An astrometric grid was established for the image using 51
stars from the USNO--A2.0 catalog (Monet et al. 1996),
with a resulting rms dispersion of $0.\!^{\prime\prime}49$.
Differences between optical and X-ray positions for
the most precisely located X-ray sources indicate that the X-ray aspect
solution agrees with the USNO--A2.0 reference frame to within
$0.\!^{\prime\prime}5$.
Additional images in $B, V, R,$ and $I$ were obtained of the
central $9^\prime \times 9^\prime$ of the same field using
the MDM 2.4m telescope on 2002 August 23--28.

Likely optical identifications of {\it Chandra} sources are listed
in Table~2, together with approximate $R$ magnitudes from
our images or from the USNO--A2.0 where available.
Only four of the {\it Chandra} sources have
no optical counterpart to a limiting magnitude $> 23$.
These happen to be the hardest sources in the image,
with $>75\%$ of their photons above 2~keV.
Thus they are likely to be AGNs that are highly absorbed by the Galactic ISM,
and not, for example, nearby, old neutron stars.
A higher-resolution image of the crowded
region around {\it Chandra} source \#2 was obtained on the MDM
2.4m telescope on 2002 November~24.  Figure~5 shows the possible
identification of that source on the 2.4m image.

\subsection{Optical Spectroscopy}

We obtained complete spectroscopic identifications for all {\it ROSAT\/}
sources within $10^\prime$ of the centroid of TeV J2032+4130, using
the Goldcam spectrograph on the KPNO 2.1m telescope
on two runs, in 2002 June and October.
The resulting spectra are shown in Figure~6.
These are sources $a$, $b$, $c$, $e$, and $f$ in Table 1, and all are bright stars.
They include one emission-line O star, one 
dMe star, and three foreground G stars, two of which are also listed in
the Massey \& Thompson (1991, hereafter MT91) compilation of stars in
Cyg OB2.  Magnitudes listed in Table~1 are from MT91, or from the
USNO--A2.0 catalog (Monet et al. 1996).

A spectrum of the $R \approx 20.4$ counterpart of {\it Chandra} source \#2,
also shown in Figure~6, was obtained with the {\it HET\/} and
Marcario Low Resolution Spectrograph on 2002 December 8.
However, its classification remains uncertain as it is
of relatively poor signal-to-noise, having been
observed through thin clouds.  All of the apparent
features in this spectrum can be attributed to imperfectly subtracted
night-sky emission lines, leaving no definite
intrinsic features in the 4600--9200 \AA\ range.

\section{A Transient X-ray Source in the Field of TeV J2032+4130}

\subsection{X-ray Properties}

None of the X-ray sources in the immediate vicinity of
TeV J2032+4130 are unusual in any way, except for 
{\it Chandra} source \#2, which lies $7^{\prime}$
from the TeV centroid.   Although this is the brightest
of the {\it Chandra} sources, with 195 photons detected,
it is noticeably absent from any of the {\it ROSAT\/}
or {\it Einstein} images.  Thus,
it may be described as a transient source.  A light curve
constructed from the 5~ks {\it Chandra} observation shows that the
source was highly variable even during this brief period (Figure~7). 
After remaining faint for the first 3.5 ks, its count rate rose
by about a factor of 10 for the final 1.5 ks.

Spectral analysis of source \#2, summarized in Table~4, is consistent with a
power law of photon index $\Gamma \sim 2.0$, or with a hot plasma of $kT \sim 6$ keV.
In addition to spectral fits for the summed observation, Table~4 presents fits for
two intervals, corresponding to the first 3438~s (low state) and the final 1478~s
(high state). 
Figure~8 shows the spectrum and best fitted power law for the summed observation.
For either spectral model,
a significant absorbing column of $N_{\rm H} \sim (1-2) \times 10^{22}$~cm$^{-2}$
is required, which is comparable to the largest X-ray and
optical extinction values measured for stars in the Cyg OB2 association
(MT91; Waldron et al. 1998).
Thus, source \#2 is likely to be either embedded in the Cyg OB2 association
at $d = 1740$~pc (MT91), or behind it.  Its peak 1--10~keV flux of
$\approx 2 \times 10^{-12}$ ergs~cm$^{-2}$~s$^{-1}$ would correspond to
a luminosity of $7 \times 10^{32}$ ergs~s$^{-1}$ at $d = 1740$~pc.
This value is somewhat larger than even the most luminous cataclysmic
variables (Eracleous, Halpern, \& Patterson 1991), and hints at a more
compact source such as a neutron star or black hole in
a quiescent low-mass X-ray binary (LMXB), or a very distant Be transient.
If so, its location could be considerably beyond Cyg OB2.

The transition in the light curve of \#2 may be an egress from an eclipse,
or, more likely for an LMXB, a ``dip''.  If so, the orbital period is probably
longer than a few hours since the eclipse or dip is at least 1 hr long.
A possible analog would be 4U~1624--49, the ``big dipper,'' which has period
of 21~hr (Watson et al. 1985; Smale, Church, \& Baluci\'nska-Church 2001).
However, assuming a distance of 8~kpc, the X-ray luminosity of \#2 would be
only $1.5 \times 10^{34}$ ergs s$^{-1}$, much less than the 
$\sim 10^{38}$ ergs s$^{-1}$ luminosity of 4U~1624--49.

\subsection{Optical and Infrared Properties}

As mentioned previously, the signal-to-noise ratio of the {\it HET\/} optical
spectrum of the counterpart to source \#2 is too poor to classify it.
However, it is likely to be the correct identification
because the X-ray and optical positions
differ by only $0.\!^{\prime\prime}37$.  The 
absence of strong TiO absorption bands rules out a late K or M dwarf,
which would otherwise be the expected spectral type for a star
of its magnitude and color located between us and the Cyg OB2
association.  Additional evidence that it is a more
distant and luminous object comes from its detection in the 2MASS survey,
with $J = 15.41 \pm 0.07,\ H = 14.17 \pm 0.06$, and $K = 13.45 \pm 0.06$. 
The absence of H$\alpha$ emission argues against a cataclysmic variable,
as does the high-state X-ray luminosity of
$\approx 7 \times 10^{32}\,(d/1.7\,{\rm kpc})^2$ ergs s$^{-1}$,
although it cannot be ruled
out that the optical spectrum is highly reddened emission from an
accretion disk in a quiescent LMXB.

If we hypothesize a very large visual extinction of $A_V = 10$, which is
compatible with the value of $N_{\rm H}$ fitted to the
X-ray spectrum, then the dereddened magnitudes
become $R = 12.9,\ J = 12.6,\ H = 12.3$, and $K = 12.3$
using the extinction curve of Cardelli, Clayton, \& Mathis (1989).
Such a flat color distribution would be compatible with accretion
disk emission in an LMXB.  Assuming a distance of 8~kpc,
the corresponding absolute magnitude $M_R =  -1.6$ is also
in the range of LMXBs (van Paradijs \& McClintock 1994).  However,
even at that large distance, the X-ray luminosity would be
only $1.5 \times 10^{34}$ ergs s$^{-1}$, and then it is not
clear that the accretion luminosity would dominate over the
secondary star in the optical.  Also, such a source would
fall far from the relation between absolute magnitude, X-ray
luminosity, and orbital period in LMXBs
delineated by van Paradijs \& McClintock (1994),
having too small an X-ray luminosity.

Alternatively, if the distance and
extinction are even larger, then it could be an early type star
such as a B star in a transient high-mass X-ray binary system.
However, the absence of H$\alpha$ emission argues against
a Be star.  Thus, there are no entirely satisfactory explanations
of source \#2 in terms of any type of X-ray binary.

\subsection{A Proton Blazar?}

While an AGN classification for the optical spectrum of source \#2 cannot
be immediately dismissed, the absence of any emission lines would favor
a BL~Lac identification, for which the lack of radio emission is highly
unusual.  Also, the known TeV blazars are highly episodic emitters, which 
is contrary to the steady nature of TeV J2032+4130.  Thus, it would not be a
simplification to hypothesize that \#2 is an AGN.  Nevertheless, it has
long been hypothesized that a class of radio-quiet blazars could exist
(Mannheim 1993; Schlickeiser 1984) that are dominated by accelerated
hadrons.  In this ``proton blazar'' model, $\gamma$-ray emission arises
from proton-induced cascades, and radio emission can be reduced if the
ratio of accelerated electrons to protons is small.  Alternatively,
an extreme blazar whose synchrotron emission peaks at MeV energies,
and inverse Compton at TeV energies, could be relevant (Ghisellini 1999).
Because HEGRA performed a sensitive survey of the Galactic plane 
(Aharonian et al. 2002b), it may not be so surprising if it turns out
that the first TeV selected blazar is discovered there.

\section{Discussion and Conclusions}

Spectroscopic optical identifications of most of the brighter
X-ray sources in $\gamma$-ray error boxes of TeV J2032+4130 and 3EG J2033+4118
are O stars in the Cyg OB2 association at $d = 1.7$ kpc, or foreground late-type
stars.  Those {\it Chandra} sources that are identified with faint optical
counterparts in the range $R \approx 17-20$, are probably M dwarfs, 
while the optically undetected sources with $R > 23$ are the
most X-ray absorbed, thus are likely background AGNs.
The only unusual X-ray source in this field
is a transient one that is the brightest source
in the recent {\it Chandra} observation.
It has a peak 1--10~keV luminosity of
$\approx 7 \times 10^{32}\,(d/1.7\,{\rm kpc})^2$ ergs s$^{-1}$.
An optical spectrum of its $R = 20.4$ possible counterpart has, albeit with
modest signal-to-noise, no strong emission or absorption features.
The hard X-ray spectrum, rapid variability, and red optical/IR colors of this object
suggest that it is a distant, quiescent X-ray binary system.   On the other
hand, it may also be the prototype of a new kind of AGN previously
hypothesized to exist, the ``proton blazar.'' If so, its X-ray flaring behavior
is a significant property, and related optical variability might be expected.

If we hypothesize that either TeV J2032+4130 or 3EG J2033+4118 is
a point source, then we have a limited number of plausible
point-source candidates at other wavelengths, perhaps only one.
Without knowing the exact nature of {\it Chandra} source \#2,
it is not a compelling identification for
either TeV J2032+4130 or 3EG J2033+4118,
especially since it lies outside both of their $2\sigma$ localization regions.
It is $7^{\prime}$ from the centroid of the TeV source, while the $2\sigma$
position uncertainty of TeV J2032+4130 is given as
$\approx 4.\!^{\prime}8$.
However, since we are faced with the first unidentified TeV source,
it is worthwhile
to pursue whatever additional observations are needed to determine
the nature of this variable X-ray candidate and to assess the
possibility of its connection with TeV J2032+4130, no matter how remote.

If TeV J2032+4130 is truly an extended source, then it need
not be centered on a point source counterpart at other wavelengths.
Benaglia et al. (2001) suggested that colliding winds from
the Cyg OB2 \#5 system and from other O stars in this association
could be responsible for the EGRET source 3EG J2033+4118.
Aharonian et al. (2002a) summarized those arguments, and hypothesized
two possible origins for extended TeV emission that may be displaced
from its originating source of energy.
One is that TeV emission could arise from $\pi^0$ decay resulting from
hadrons accelerated in shocked OB star winds and interacting with a local,
dense gas cloud.  The other is inverse Compton TeV emission
in a jet-driven termination shock, either from an as-yet undetected microquasar,
or from Cyg X-3.  Another reason to investigate the nature
of {\it Chandra} \#2 would be to find out if it could be such
a jet source.  We plan to pursue more detailed X-ray and optical studies
of this source.

\acknowledgements

This publication makes use of data obtained from HEASARC at Goddard Space Flight Center 
and the SIMBAD astronomical database.  It also
makes use of data products from the Two Micron All Sky Survey,
which is a joint project of the University of Massachusetts and the Infrared Processing 
and Analysis Center/California Institute of Technology, funded by the National
Aeronautics and Space Administration and the National Science Foundation.
This work was based in part on observations obtained with the Hobby-Eberly
Telescope, which is a joint project of the University of Texas at Austin,
Pennsylvania State University, Stanford University,
Ludwig-Maximillians-Universit\"at M\"unchen, and Georg-August-Universit\"at
G\"ottingen.  The Marcario Low-Resolution Spectrograph is a joint project
of the Hobby-Eberly Telescope Partnership and the Instituto de Astronomia
de la Universidad Nacional Aut\'onoma de Mexico.
R. M. acknowledges support from NSF grant PHY-9983836.
E.V.G. is supported by NASA LTSA grant NAG~5-7935.
{\it Chandra} studies of unidentified $\gamma$-ray sources is supported
by SAO grants GO2-3071X and GO2-3082X to J.P.H.

\clearpage

\begin{deluxetable}{ccccllccc}
\rotate
\tablenum{1}
\tablecolumns{9}
\tablewidth{0pc}
\tablecaption{{\it ROSAT\/} Sources in the Field of 3EG J2033+4118 and TeV J2032+4130}
\tablehead
{
{\it ROSAT} & {\it Chandra} &  X-ray Position & Optical Position & Name & Spectral & $B$ & $V$ & $R$\\
Ident. & cross ref.   & R.A.\tablenotemark{a}\hskip 3em Decl.\tablenotemark{b} & R.A.\tablenotemark{a}\hskip 3em Decl.\tablenotemark{b} &  & Type & (mag) & (mag) & (mag)
}
\startdata
 $a$ & . . .   & 20 32 18.1 +41 28 07 &  20 32 19.244 +41 27 57.37 &  . . .      & G8V        & 15.4  & . . . & 15.0 \\
 $b$ & . . .   & 20 32 41.5 +41 27 44 &  20 32 41.454 +41 27 43.51 &  . . .      & M5Ve       & 16.8  & . . . & 15.8 \\
 $c$ & 18      & 20 32 16.1 +41 27 08 &  20 32 13.836 +41 27 12.33 & Cyg OB2 \#4 & O7III((f)) & 11.42 & 10.23 & 10.2 \\
 $d$ & . . .   & 20 33 03.9 +41 24 14 &  20 33 03.902 +41 24 08.48 &  . . .      & . . .      & 16.4  & . . . & 16.9 \\
 $e$ & 25      & 20 31 36.5 +41 23 38 &  20 31 37.267 +41 23 36.01 & MT91 \#115  & G6V        & 13.90 & 13.10 & 13.1 \\
 $f$ & 26      & 20 31 51.5 +41 23 21 &  20 31 51.319 +41 23 23.79 & MT91 \#152  & G3V        & 13.40 & 12.74 & 13.1 \\
 $g$ &  . . .  & 20 33 14.0 +41 20 14 &  20 33 14.144 +41 20 22.00 & Cyg OB2 \#7 & O3If     & 11.94 & 10.50 & . . .\\
 $h$ &  . . .  & 20 33 15.3 +41 18 47 &  20 33 15.077 +41 18 50.50 & Cyg OB2 \#8A & O5.5I(f)       & 10.08 &  8.99 &  9.1 \\
 $i$ &  . . .  & 20 32 22.7 +41 18 17 &  20 32 22.432 +41 18 18.85 & Cyg OB2 \#5 & O7e         & 10.64 &  9.21 &  8.1 \\
 $j$ & . . .   & 20 33 11.0 +41 15 07 &  20 33 10.733 +41 15 08.22 & Cyg OB2 \#9 & O5Iab:e    & 12.61 & 10.78 & . . .\\
 $k$ &  . . .  & 20 32 41.4 +41 14 28 &  20 32 40.957 +41 14 29.30 & Cyg OB2 \#12 & B5Iab:    & 14.41 & 11.40 & . . .\\
 $l$ &  . . .  & 20 32 32.1 +41 14 11 &  20 32 31.556 +41 14 08.48 & MT91 \#267  &  . . .     & 15.06 & 12.87 & 11.8 \\
 $m$ & 27      & 20 31 38.0 +41 13 21 &  20 31 37.504 +41 13 21.05 & Cyg OB2 \#3 & O9:         & 11.50 & 10.35 &  9.3 \\
 $n$ &  . . .  & 20 33 09.4 +41 13 21 &  20 33 08.820 +41 13 18.00 & MT91 \#417 & O4III(f)   & 13.59 & 11.5  &  9.3 \\
 $o$ &  . . .  & 20 33 02.3 +41 11 19 &  20 33 01.793 +41 11 11.41 & MT91 \#384  &  . . .     & 17.28 & 16.73 & 16.1 \\
 $p$ &  . . .  & 20 33 23.3 +41 09 07 &  20 33 23.477 +41 09 13.38 & MT91 \#516  & O5.5V((f)) & 14.04 & 11.84 & 11.0 \\
 $q$ &  . . .  & 20 32 07.9 +41 08 37 &  20 32 07.330 +41 08 50.56 & . . .       & . . .      & 14.7  & 14.2  & 13.6  \\
 $r$ &  . . .  & 20 33 41.6 +41 08 01 &  20 33 42.036 +41 07 53.70 & MT91 \#615  &  . . .     & 12.2  & 11.51 & 11.2 \\
\enddata
\tablenotetext{a}{Units of right ascension are hours, minutes, and seconds.}
\tablenotetext{b}{Units of declination are degrees, arcminutes, and arcseconds.}
\end{deluxetable}

\begin{deluxetable}{cccccclc}
\rotate
\tablenum{2}
\tablecolumns{8}
\tablewidth{0pc}
\tablecaption{{\it Chandra} Sources in the Field of TeV J2032+4130}
\tablehead
{
{\it Chandra} & {\it ROSAT\/} &  X-ray Position\tablenotemark{a} & Counts\tablenotemark{b}
& Counts\tablenotemark{b} & Optical Position\tablenotemark{a} & Name & $R$ \\
Ident. & cross ref. & R.A.\hskip 4em Decl. & ($< 2$ keV) & ($> 2$ keV) & R.A.\hskip 4em Decl. & & (mag)
}
\startdata
 1 &  . . .  & 20 31 56.426 +41 37 23.35 &  7 & 35 &           . . .            & . . .  & $>23.2$  \\
 2 &  . . .  & 20 31 43.755 +41 35 55.17 & 77 &118 & 20 31 43.739 +41 35 55.49  & . . .  & 20.4  \\
 3 &  . . .  & 20 32 10.248 +41 35 10.21 &  4 & 10 & 20 32 10.219 +41 35 11.31  & . . .  & 18.3  \\
 4 &  . . .  & 20 32 25.392 +41 34 01.92 & 16 &  4 & 20 32 25.346 +41 34 02.10  & MT91 \#249  & 12.2 \\ 
 5 &  . . .  & 20 32 05.328 +41 33 12.38 &  3 & 10 &         . . .              & . . .  & $>23.7$  \\
 6 &  . . .  & 20 32 27.456 +41 32 50.78 &  3 & 10 &         . . .              & . . .  & $>23.7$  \\
 7 &  . . .  & 20 31 51.855 +41 31 18.91 &  3 & 22 &         . . .              & . . .  & $>23.7$  \\
 8 &  . . .  & 20 32 30.504 +41 31 01.74 &  6 &  5 & 20 32 30.412 +41 31 02.88  & . . .  & 21.2 \\
 9 &  . . .  & 20 31 39.192 +41 30 18.00 &  9 &  2 & 20 31 39.104 +41 30 18.49  & . . .  & 17.1 \\
10 &  . . .  & 20 31 23.544 +41 29 48.77 & 13 &  9 & 20 31 23.573 +41 29 49.45  & . . .  & 15.1 \\
11 &  . . .  & 20 32 18.845 +41 29 32.42 & 20 & 15 & 20 32 18.812 +41 29 32.83  & . . .  & 19.3 \\
12 &  . . .  & 20 32 37.848 +41 28 52.39 & 10 &  9 & 20 32 37.831 +41 28 52.93  & . . .  & 16.7:\tablenotemark{c}\\
13 &  . . .  & 20 32 25.752 +41 28 42.53 &  8 &  7 & 20 32 25.731 +41 28 42.89  & . . .  & 17.3 \\
14 &  . . .  & 20 32 28.080 +41 28 28.52 &  3 & 19 & 20 32 28.027 +41 28 29.02  & . . .  & 18.8 \\
15 &  . . .  & 20 32 52.163 +41 28 15.92 & 10 & 14 & 20 32 52.206 +41 28 17.17  & . . .  & 19.3:\tablenotemark{c}  \\
16 &  . . .  & 20 32 42.072 +41 27 48.13 & 17 &  1 & 20 32 42.014 +41 27 48.32  & . . .  & 15.7 \\
17 &  . . .  & 20 32 14.688 +41 27 39.67 &  7 &  8 & 20 32 14.693 +41 27 40.09  & MT91 \#221  & 12.0 \\
18 &  $c$    & 20 32 13.843 +41 27 12.13 & 27 &  8 & 20 32 13.837 +41 27 12.34  & Cyg OB2 \#4 & 10.2 \\
19 &  . . .  & 20 31 33.662 +41 26 51.46 & 15 & 22 & 20 31 33.650 +41 26 51.71  & . . .  & 19.6 \\
20 &  . . .  & 20 32 27.598 +41 26 21.62 & 20 &  3 & 20 32 27.663 +41 26 22.44  & MT91 \#258  & 10.4 \\
21 &  . . .  & 20 32 45.432 +41 25 36.44 & 19 &  4 & 20 32 45.462 +41 25 37.51  & Cyg OB2 \#6 & 10.1 \\
22 &  . . .  & 20 32 55.344 +41 25 17.00 & 15 &  8 & 20 32 54.773 +41 25 16.09  & MT91 \#360  & 16.2 \\
23 &  . . .  & 20 32 06.877 +41 25 10.61 & 23 & 26 & 20 32 06.826 +41 25 10.68  & . . .  & 20.3 \\
24 &  . . .  & 20 31 47.568 +41 24 48.02 & 12 & 15 & 20 31 47.535 +41 24 48.43  & . . .  & 18.1 \\
25 &  $e$    & 20 31 37.251 +41 23 35.44 & 29 &  3 & 20 31 37.267 +41 23 36.01  & MT91 \#115  & 13.1 \\
26 &  $f$    & 20 31 51.321 +41 23 22.75 & 54 &  7 & 20 31 51.319 +41 23 23.79  & MT91 \#152  & 13.1 \\
27 &  $m$    & 20 31 37.156 +41 13 17.42 &169 & 59 & 20 31 37.504 +41 13 21.05  & Cyg OB2 \#3 &  9.3 \\
\enddata
\tablenotetext{a}{Units of right ascension are hours, minutes, and seconds.
Units of declination are degrees, arcminutes, and arcseconds.}
\tablenotetext{b}{Soft and hard counts in a $12^{\prime\prime}$ radius aperture.
The total included background is estimated as 1--3 counts.}
\tablenotetext{c}{Crowded field.}
\end{deluxetable}

\begin{deluxetable}{ccr}
\tablenum{3}
\tablecolumns{3}
\tablewidth{0pc}
\tablecaption{NVSS Sources in the Field of TeV J2032+4130}
\tablehead{
R.A.\tablenotemark{a}   &  Decl.\tablenotemark{b}  &  $F_{1.4 \rm GHz}$ \\
& &  (mJy)
}
\startdata
20 31 31.18 & +41 25 12.6 &  2.1 \\
20 31 45.08 & +41 35 34.2 & 18.1 \\
20 31 48.17 & +41 33 58.3 &  6.1 \\
20 31 56.81 & +41 35 31.4 &  4.5 \\
20 32 01.34 & +41 37 22.7 & 14.0 \\
20 32 08.33 & +41 29 17.4 &  3.8 \\
20 32 17.04 & +41 26 21.4 &  2.3 \\
20 32 52.27 & +41 30 09.5 & 10.6 \\
\enddata
\tablenotetext{a}{Units of right ascension are hours, minutes, and seconds.}
\tablenotetext{b}{Units of declination are degrees, arcminutes, and arcseconds.}
\end{deluxetable}

\begin{deluxetable}{lccc}
\tablenum{4}
\tablecolumns{4}
\tablewidth{0pc}
\tablecaption{Spectral Fits to {\it Chandra} Source \#2}
\tablehead{
Parameter  &  Total Interval & Faint Interval & Bright Interval
}
\startdata
Counts (1--6 keV)        & 169   &    35  &   134   \\
Exposure Time (s)        & 4916  &  3438  &  1478   \\
Count Rate (s$^{-1}$)    & 0.03  &  0.01  &  0.09   \\
\hline
Power Law:               &       &        &         \\
\hline
$N_{\rm H}\ (10^{22}$ cm$^{-2}$) &  $1.5 (0.6-2.3)$\tablenotemark{a} &  1.3  &  1.7  \\
$\Gamma$                         &  $2.0 (1.2-2.7)$\tablenotemark{a} &  2.1  &  2.0  \\
Flux ($10^{-13}$ ergs cm$^{-2}$ s$^{-1}$)\tablenotemark{b}   &  6.6  &  1.8  & 18.0  \\
$\chi^2_{\nu}$(dof)      & 0.6(13) & 1.0(2) & 0.4(10) \\
\hline
Raymond-Smith:                            &       &       &       \\
\hline
$N_{\rm H}\ (10^{22}$ cm$^{-2}$)  &  $1.2 (0.6-1.8)$\tablenotemark{a}  &  0.9  &  1.6  \\
$kT$ (keV)                       &  $6.0 (3.0-\infty)$\tablenotemark{a}  &  6.6  &  4.0  \\
Flux ($10^{-13}$ ergs cm$^{-2}$ s$^{-1}$)\tablenotemark{b}  &  7.1 & 2.1 &  17.4 \\
$\chi^2_{\nu}$(dof)     & 0.6(13)  & 1.1(2) & 0.3(10) \\
\enddata
\tablenotetext{a}{$1\sigma$ uncertainty range for two interesting parameters.}
\tablenotetext{b}{Absorbed flux in the 1--10 keV energy band.}
\end{deluxetable}

\newpage

\begin{figure}[t]
  \begin{center}
    \includegraphics[height=40pc]{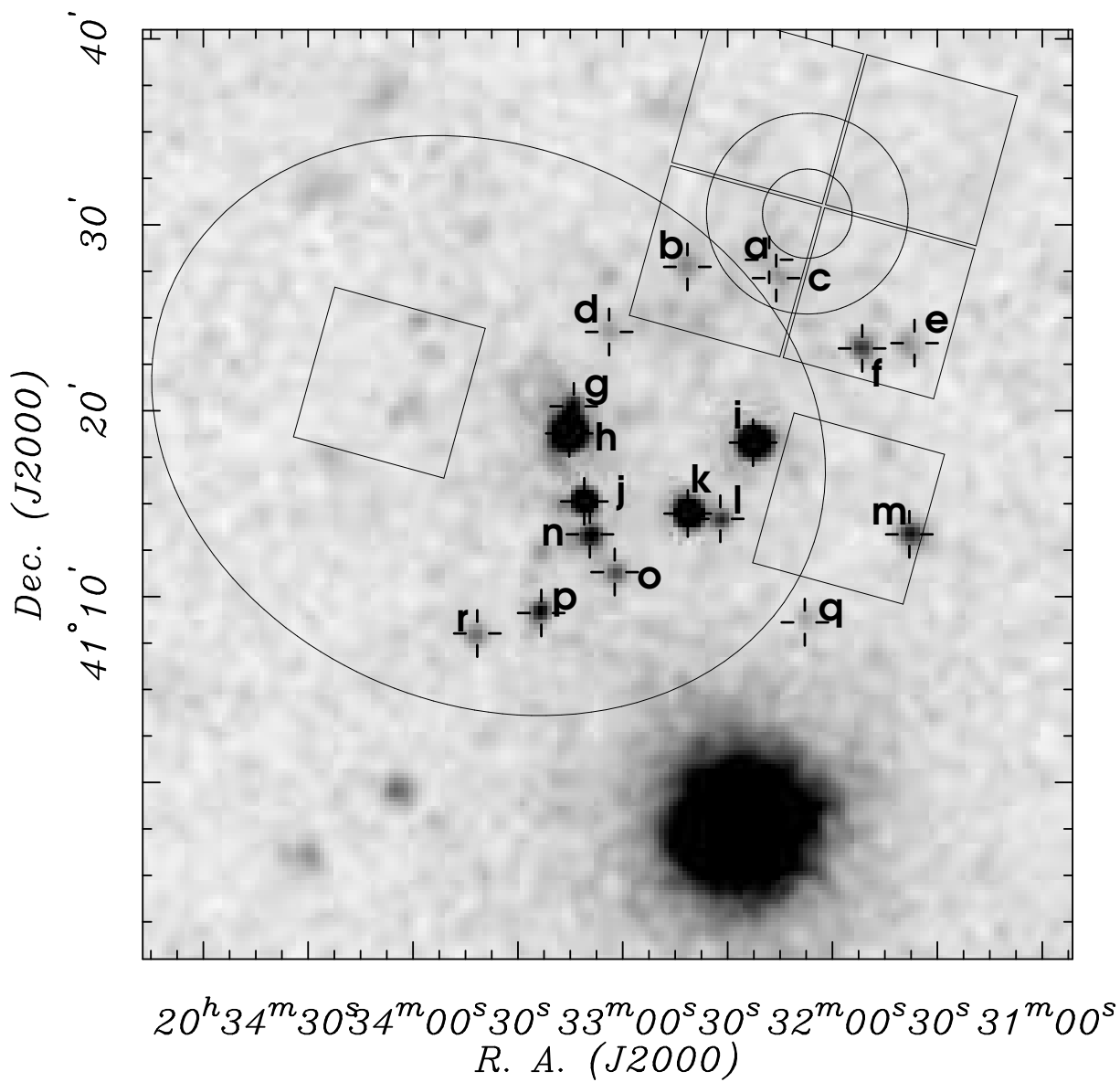}
  \end{center}
  \caption{{\sl ROSAT\/} PSPC X-ray image.  The bright source is Cyg X-3.
The properties of the numbered sources are given in Table 1.
The ellipse is the 95\% uncertainty location of 3EG J2033+4118
from Mattox et al. (2001).  The small circle is the $1\sigma$ uncertainty of
the centroid of TeV J2032+4130, and the large circle
is the estimated Gaussian $1\sigma$ extent of the TeV emission (Aharonian et al. 2002a).
The squares are the fields of view of the CCDs in the subsequent
{\it Chandra} observation (see Figure~4).}
\label{fig1}
\end{figure}

\newpage

\begin{figure}[t]
  \begin{center}
    \includegraphics[height=20pc]{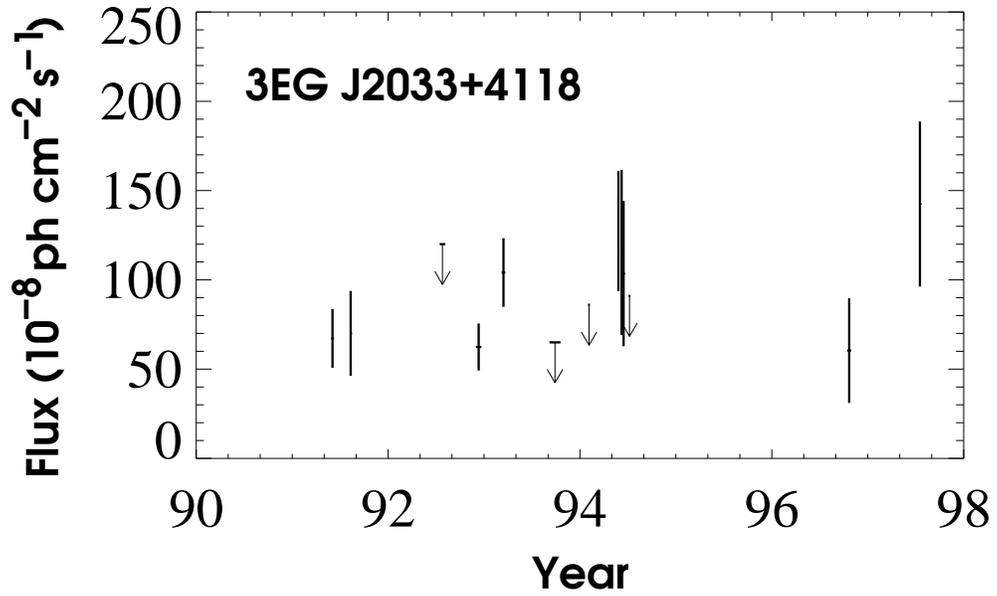}
  \end{center}
  \caption{EGRET $\gamma$-ray light curve for 3EG J2033+4118, 1991--1995
from Hartman et al. (1999), and 1996--1997 from this paper.
Arrows are $2 \sigma$ upper limits. The horizontal
error bars correspond to the extent of an individual observation.}
\label{fig2}
\end{figure}

\newpage
\begin{figure}[t]
  \begin{center}
    \includegraphics[height=35pc,angle=-90]{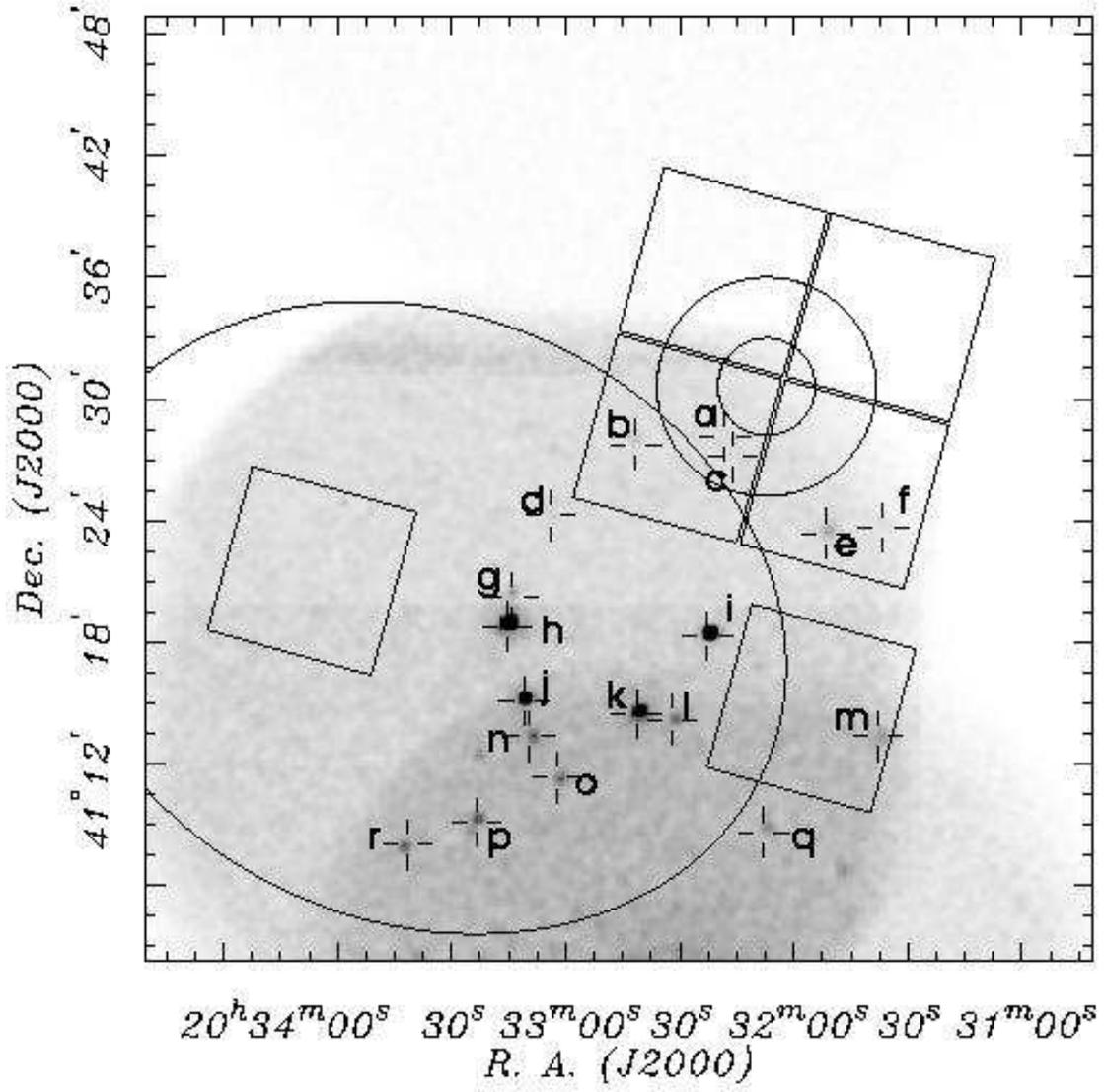}
  \end{center}
  \caption{Same as Figure~1, but for the {\it ROSAT\/} HRI X-ray images
in the field of 3EG J2033+4118 and TeV J2032+4130.}
\label{fig3}
\end{figure}

\newpage

\begin{figure}[t]
  \begin{center}
    \includegraphics[height=40pc,angle=-90]{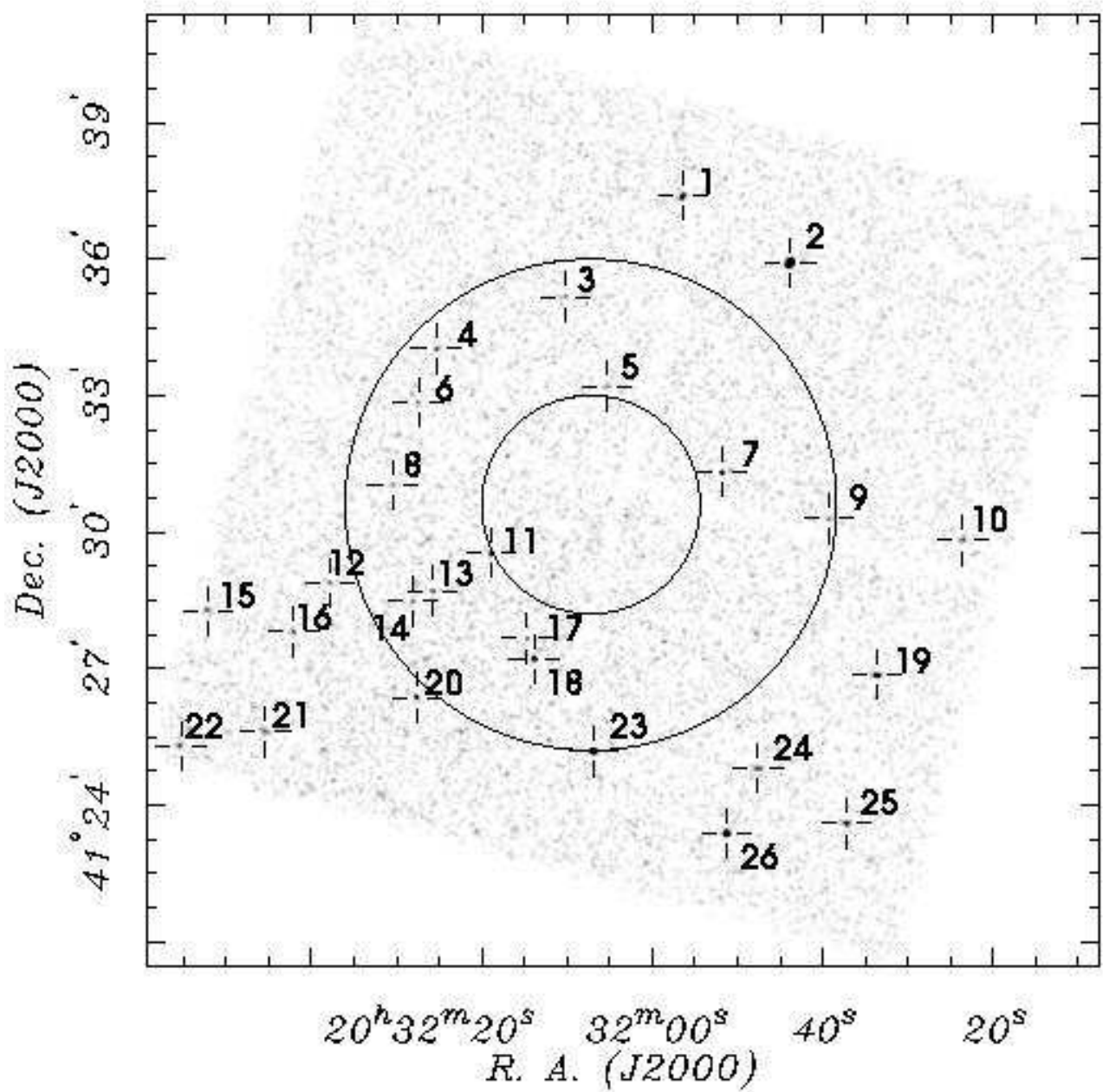}
  \end{center}
\vspace{-0.5cm}
  \caption{{\it Chandra} ACIS-I image of the field of TeV J2032+4130.
The properties of the
identified sources are given in
in Table 2. The small circle is the $1\sigma$ uncertainty of
the centroid of TeV J2032+4130, and the large circle
is the estimated Gaussian $1\sigma$ extent of the TeV emission (Aharonian et al. 2002a).
The brightest {\it Chandra} source \#2 was not detected in {\it ROSAT\/} images.}
\label{fig4}
\end{figure}

\newpage

\begin{figure}[t]
  \begin{center}
    \includegraphics[height=40pc,angle=-90]{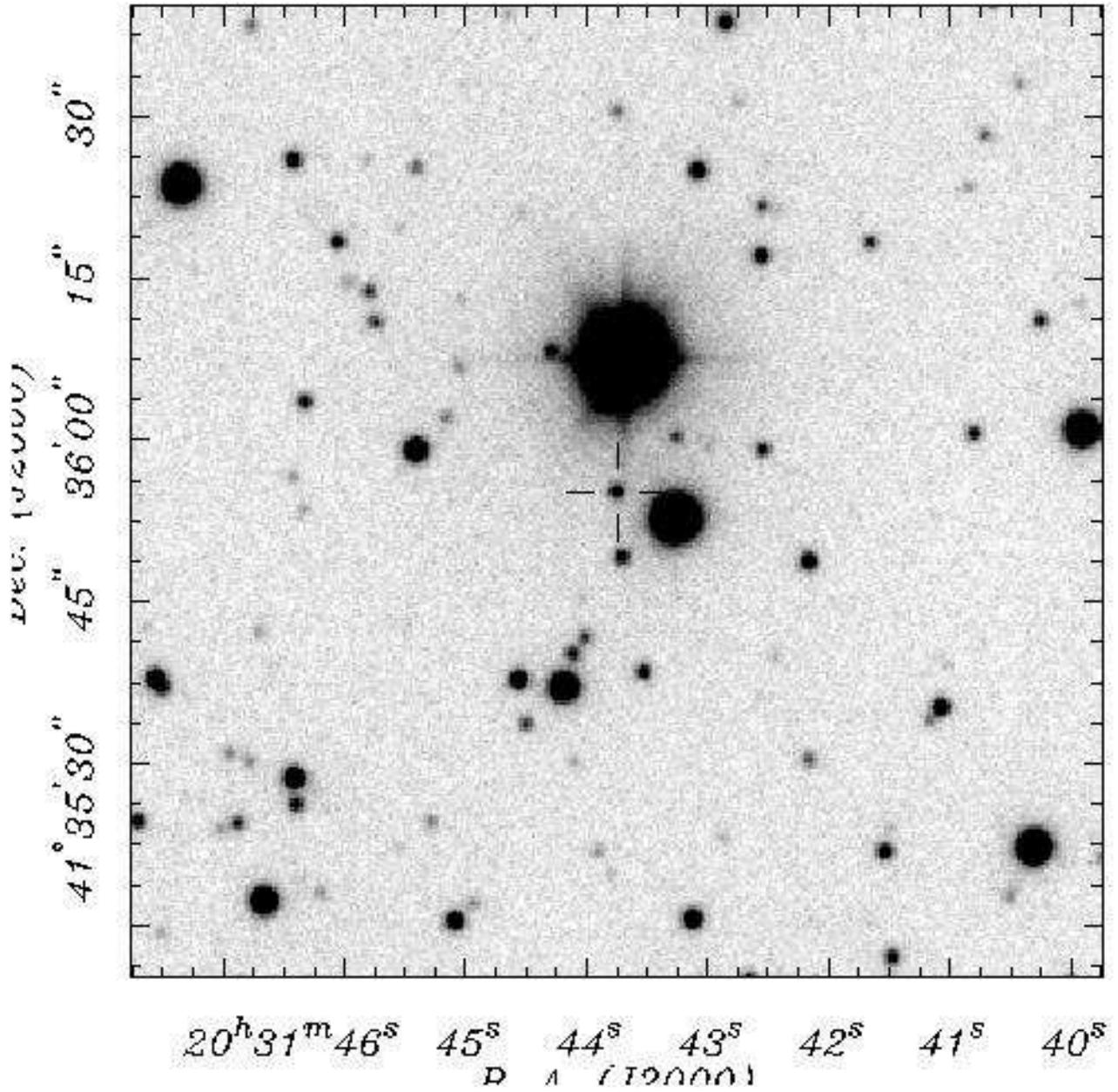}
  \end{center}
  \caption{$R$-band finding chart for {\it Chandra\/} source \#2 taken with MDM 2.4m
telescope on 2002 November 24.  Seeing is $1.\!^{\prime\prime}0$. The {\it cross}
marks the X-ray position, which is coincident with a star of magnitude $R = 20.4$.}
\label{fig5}
\vspace{-0.5cm}
\end{figure}

\newpage

\begin{figure}[t]
  \begin{center}
    \includegraphics[height=40pc]{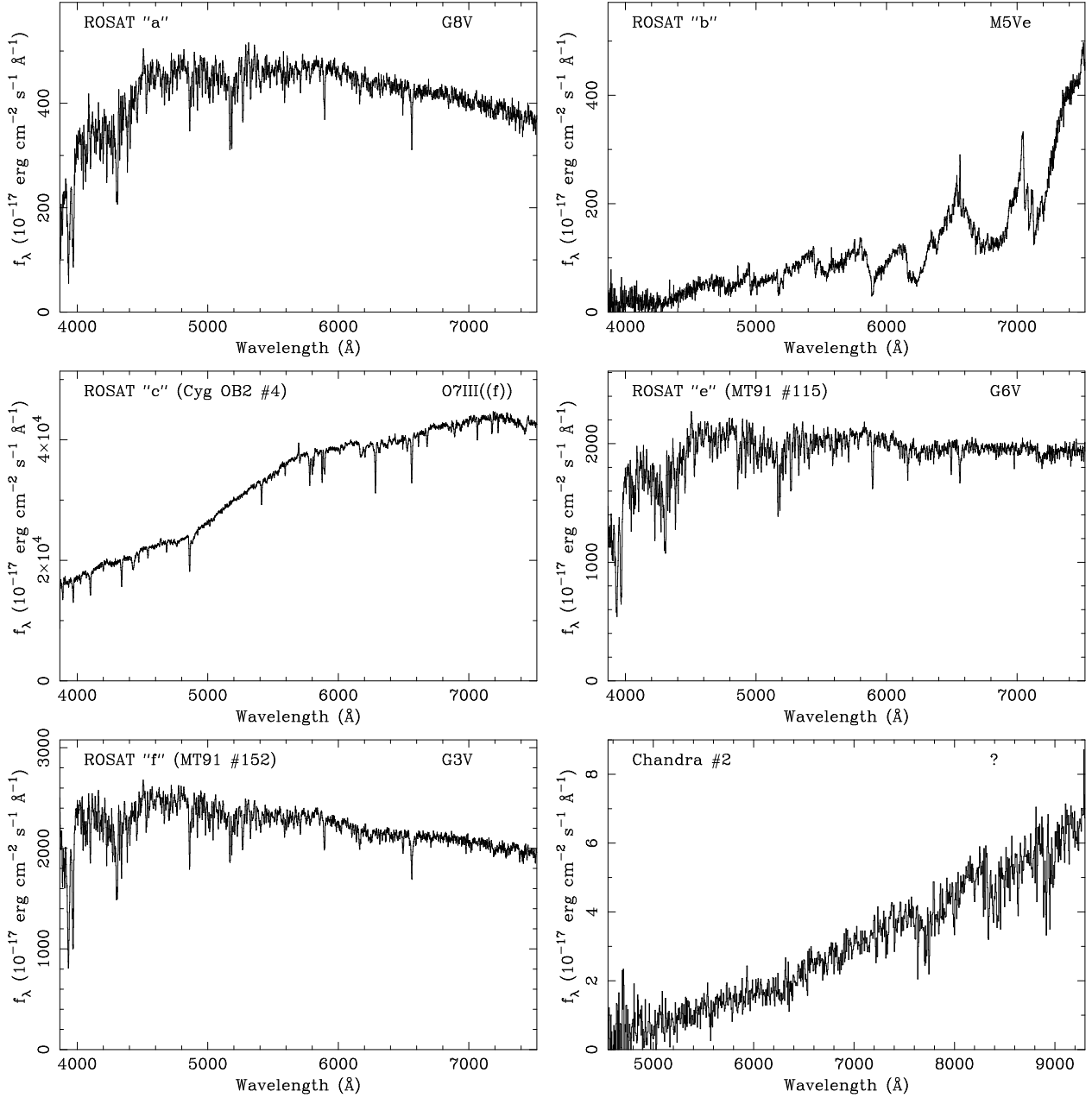}
  \end{center}
  \caption{Optical spectra of five {\it ROSAT\/} sources in the field of TeV J2032+4130
taken with the KPNO 2.1m telescope, and one {\it Chandra} source with the {\it HET}.}
\label{fig6}
\end{figure}

\newpage

\begin{figure}[t]
  \begin{center}
    \includegraphics[height=40pc]{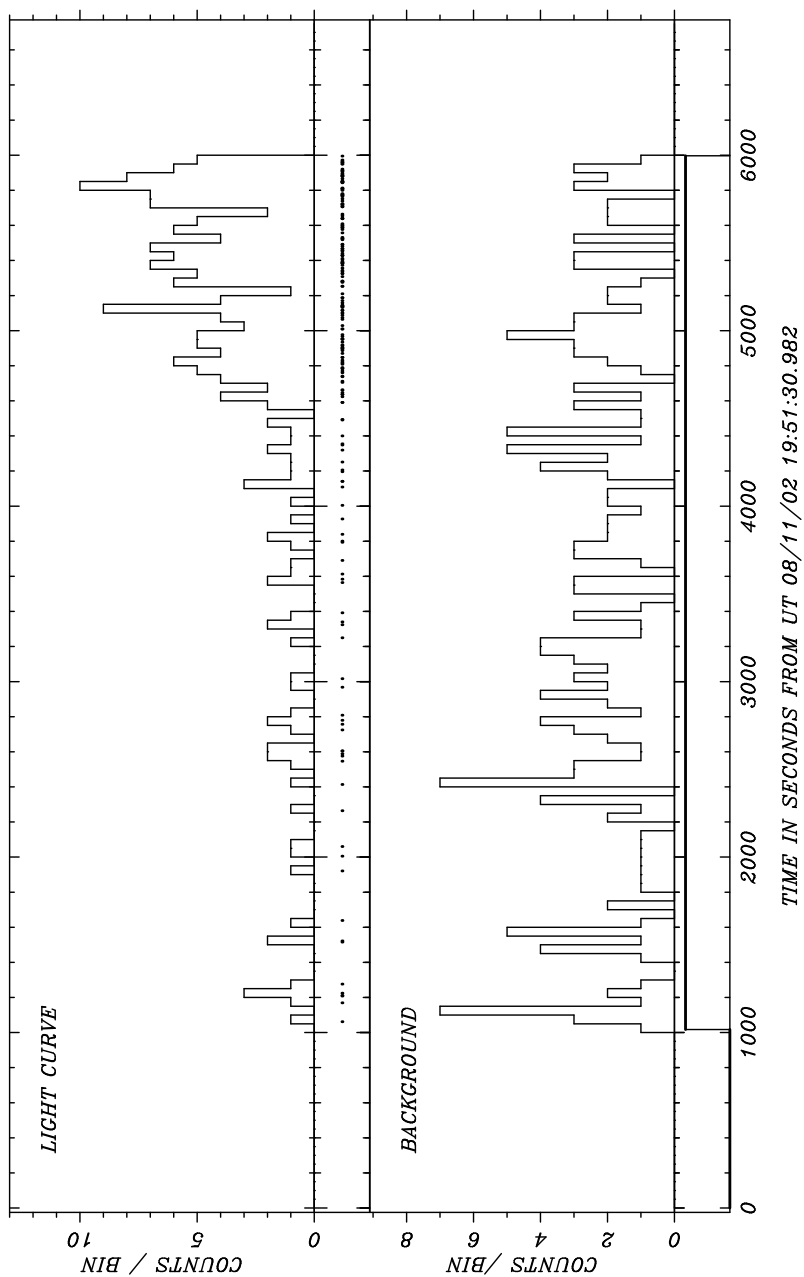}
  \end{center}
  \caption{{\it Upper panel}: Light curve of {\it Chandra\ } source \#2 
extracted from an aperture of radius $12^{\prime\prime}$, in 50 s time bins.
The {\it dots} below the light curve are the arrival times of the individual photons.
{\it Lower panel}: Local background extracted from an area 33
times larger than the source extraction region.  Thus, background is
demonstrated to be stable, and a negligible contaminant of the source light curve.}
\label{fig7}
\vspace{1cm}
\end{figure}

\newpage

\begin{figure}[t]
  \begin{center}
   \includegraphics[height=40pc]{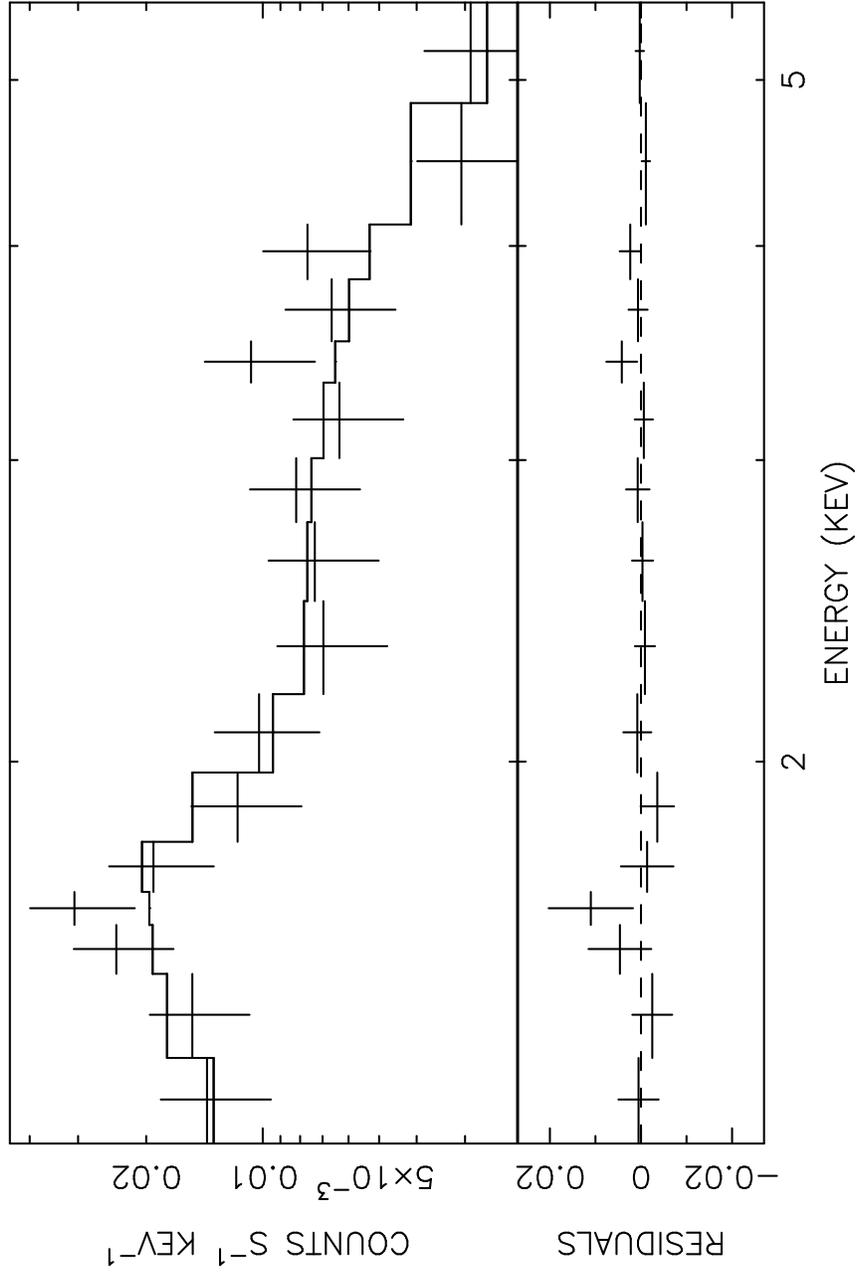}
  \end{center}
  \caption{Spectrum of {\it Chandra} source \#2 ({\it crosses}) and best fitted
power-law model ({\it solid line}).
This spectrum is for the entire time interval extracted from a
12$^{\prime\prime}$ radius aperture.  The background is treated as negligible,
as demonstrated in Figure~7.}
\label{fig8}
\end{figure}

\end{document}